\documentclass[aps,prd,twocolumn,superscriptaddress]{revtex4}
\usepackage{epsfig,epsf}
\usepackage{amsmath}
\usepackage{amsthm}
\usepackage{amsfonts}
\usepackage{amssymb}
\usepackage{dsfont}
\usepackage{multirow}
\usepackage{appendix}
\usepackage{slashed}
\usepackage[active]{srcltx}
\usepackage{psfrag}

\begin{document}

\title{Open charm-bottom axial-vector tetraquarks and their properties}
\date{\today}
\author{S.~S.~Agaev}
\affiliation{Institute for Physical Problems, Baku State University, Az--1148 Baku,
Azerbaijan}
\author{K.~Azizi}
\affiliation{Department of Physics, Do\v{g}u\c{s} University, Acibadem-Kadik\"{o}y, 34722
Istanbul, Turkey}
\author{H.~Sundu}
\affiliation{Department of Physics, Kocaeli University, 41380 Izmit, Turkey}

\begin{abstract}
The charged axial-vector $J^{P}=1^{+}$ tetraquarks $Z_{q}=[cq][\bar {b} \bar
q ]$ and $Z_{s}=[cs][\bar {b} \bar s]$ with the open charm-bottom contents
are studied in the diquark-antidiquark model. The masses and meson-current
couplings of these states are calculated by employing QCD two-point sum rule
approach, where the quark, gluon and mixed condensates up to eight
dimensions are taken into account. These parameters of the tetraquark states $%
Z_{q}$ and $Z_{s}$ are used to analyze the vertices $Z_q B_c \rho $ and $Z_s
B_c \phi $ to determine the strong $g_{Z_qB_c \rho }$ and $g_{Z_sB_c \phi }$
couplings. For these purposes, QCD light-cone sum rule method and its
soft-meson approximation are utilized. The couplings $g_{Z_qB_c \rho }$ and $%
g_{Z_sB_c \phi }$, extracted from this analysis, are applied for evaluating
of the strong $Z_q \to B_c \rho$ and $Z_s \to B_c \phi$ decays' widths,
which are essential results of the present investigation. Our predictions
for the masses of the $Z_{q}$ and $Z_{s}$ states are confronted with similar
results available in the literature.
\end{abstract}

\maketitle

\section{Introduction}

\label{sec:Int}

Charmonium-like states discovered during last years mainly in the exclusive
B-meson decays as resonances in the relevant mass distributions became
interesting objects for both experimental and theoretical studies in high
energy physics. Conventional hadrons, composed of two and three quarks, and
investigated in a rather detailed form constitute main part of the known
particles. At the same time, the theory of the strong interactions -- the
Quantum Chromodynamics does not contain principles excluding an existence of
the multiquark states. The tetraquark and pentaquark states composed of the
four and five valence quarks, respectively, and hybrids built of the quarks
and gluons are among most promising candidates to occupy the vacant shelves
in the multiquark spectroscopy. Due to joint efforts of experimentalists and
theorists considerable progress in understanding of the quark-gluon
structure of the multiquark --exotic states and explaining of their
properties were achieved, but remaining questions are more numerous that
answered ones (for latest reviews, see Refs.\ \cite%
{Chen:2016qju,Chen:2016spr,Esposito:2014rxa,Meyer:2015eta}).

The main source of problems, which complicates the studying of the
charmonium-like tetraquarks, is the existence of the conventional charmonium
states in the energy ranges of the exploring decay processes. Charmonia
generate difficulties in interpretation of experimental results, because the
pure $c\bar{c}$ states may emerge as the resonances in the mass
distributions of the processes, or generate background effects due to states
dynamically connected with $c\bar{c}$ levels. Only after eliminating effects
of the charmonium states in forming of the experimental data, observed
resonances can be considered as real exotic particles. The well-known $%
X(3872)$ state is the best sample to illustrate existing problems. It was
discovered as very narrow resonance in B meson decay $B\to KX \to KJ/\psi
\rho \to KJ/\psi\pi^{+} \pi^{-}$ by the Belle Collaboration \cite{Belle:2003}%
, and was later confirmed in CDF, D0 and BaBar experiments (see, Refs.\ \cite%
{CDF:2004,D0:2004,BaBar:2005}). Its other production mechanisms running
through decay chains $B\to KX \to KJ/\psi\omega \to KJ/\psi\pi^{+}
\pi^{-}\pi^{0}$, $B\to KX \to KJ/\psi \gamma$ and $B\to KX \to K\psi(2S)
\gamma$ were also experimentally measured and comprehensively studied \cite%
{Abe:2005ix,Aubert:2008ae}. The gathered information poses severe
restrictions on theoretical models claiming to describe a behavior of the $%
X(3872)$ state. Attempts were made to explain the collected data by treating
$X(3872)$ as the excited conventional charmonium $\chi_{c1}(2^{3}P_{1})$ \cite%
{Barnes:2005pb}, or as the state formed due to dynamical coupled-channel
effects \cite{Danilkin:2010cc}. It was considered in the context of
four-quark compounds, both as the $D\bar{D}^{\star}$ molecule or its
admixtures with the charmonium states \cite%
{Close:2003sg,Tornqvist:2004qy,Zanetti:2011ju,Guo:2014taa}, and as the
diquark-antidiquarks states \cite%
{Maiani:2004vq,Maiani:2007vr,Navarra:2006nd,Dubnicka:2010kz,Wang:2013vex}.

But existence of the tetraquarks, which do not contain $\bar{c}c$ or $\bar{b}%
b$ pairs is also possible, because fundamental laws of QCD do not forbid
production of such resonances in hadronic processes. These particles may
appear in the exclusive reactions as the open charm (i.e., as states
containing $c$ or $\bar{c}$ quarks) and open bottom resonances. The $%
D_{s0}^{\star}(2317)$ and $D_{s1}(2460)$ mesons discovered by the BaBar and
CLEO collaborations \cite{Aubert:2003fg,Besson:2003cp}, are now being
considered as candidates to open charm tetraquark states. The $X(5568)$
resonance remains a unique candidate to the open bottom tetraquark, which is
also  a particle containing four different quarks.
Unfortunately, the experimental situation formed around $X(5568)$ remains
unclear. Indeed, the evidence for $X(5568)$ was first announced by the D0
Collaboration in Ref.\ \cite{D0:2016mwd}. Later it was seen again  by D0 in
the $B_{s}^0$ meson's semileptonic decays \cite{D0}. Nevertheless, the LHCb
and CMS collaborations could not see the same resonance from analysis of
their experimental data \cite{Aaij:2016iev,CMS:2016}. Theoretical
investigations aiming to explain the nature of $X(5568)$ and calculate its
parameters lead also to contradictory conclusions. Predictions obtained in
some of these works are in a nice agreement with results of the D0
Collaboration, while in others even an existence of the $X(5568)$ state is
an object of doubts. The detailed discussions of these and related questions
of the $X(5568) $ state' physics can be found in original works (see,
Ref.\ \cite{Chen:2016spr} and references therein).

The open charm-bottom tetraquarks belong to  another type of exotic
states. They already attracted an interest of physicists even till now were not observed
experimentally. The original investigations of these particles
started more than two decades ago, and, therefore, a considerable
theoretical information on their expecting properties is available in the
literature. For example,
the open charm-bottom type tetraquarks with the
contents $\{Qq\}\{Q^{\prime}q\}$, $\{Qs\}\{Q^{\prime}s\}$ and molecule
structures were considered in Refs.\ \cite{Zhang:2009vs} and \cite%
{Zhang:2009em}, respectively. In these papers the masses of these
hypothetical states were calculated in the context of QCD two-point sum rule
approach using in the operator product expansion (OPE) the operators up to
dimension six. In the framework of the diquark-antidiquark model the open
charm-bottom states were analyzed in Ref.\ \cite{Chen:2013aba}. In order to
extract masses of these states, the authors again utilized QCD sum rule
method and interpolating currents of different color structure. Other
aspects of these tetraquark systems can be found in Refs.\ \cite%
{Zouzou:1986qh,SilvestreBrac:1993ry,Ebert:2007rn,Sun:2012sy,Albuquerque:2012rq}.

In a previous article \cite{Agaev:2016dsg} we explored the charged scalar $J^{P}=0^{+}$
tetraquark states $Z_{q}=[cq][\bar {b}\bar q ]$ and $Z_{s}=[cs][\bar {b}\bar
s]$ in the context of the diquark-antidiquark model, and calculated their masses and
widths some of their decay channels. In the present work we extend our investigations by
including into analysis the axial-vector $J^{P}=1^{+}$ $Z_{q}=[cq][\bar{b}\bar{q}]$
and $Z_{s}=[cs][\bar{b}\bar{s}]$ open charm-bottom tetraquarks, and their kinematically
allowed decay modes.

We start from calculation of their masses and meson-current couplings. For
these purposes, we employ QCD two-point sum rule method, which was invented
to calculate parameters of the conventional hadrons \cite{Shifman:1979}, but
soon was applied to analysis of the exotic states, as well (see, Refs.\ \cite%
{Braun:1985ah,Braun:1988kv,Balitsky:1982ps,Reinders:1985}). The parameters of the
open charm-bottom tetraquarks obtained within this method are used to explore the
strong vertices $Z_{q}B_{c}\rho$ and $Z_{s}B_{c}\phi$, and calculate the
corresponding couplings $g_{Z_{q}B_{c}\rho}$ and $g_{Z_{s}B_{c}\phi }$.
These couplings are required to evaluate the widths of the $Z_{q}\rightarrow
B_{c}\rho $ and $Z_{s}\rightarrow B_{c}\phi $ decays. To this end, we apply
QCD light-cone sum method and soft-meson approximation proposed in Refs.\
\cite{Braun:1989,Ioffe:1983ju,Braun:1995}. For analysis of the strong
vertices of tetraquarks the method was, for the first time, examined in
Ref.\ \cite{Agaev:2016dev}, and afterwards successfully used to investigate
decay channels of some  tetraquarks states (see, Refs.\ \cite%
{Agaev:2016ijz,Agaev:2016lkl,Agaev:2016urs}).

The present work is organized in the following manner. In Sec.\ \ref%
{sec:Mass} we calculate the masses and meson-current couplings of the
axial-vector tetraquarks with open charm-bottom contents. Section \ref%
{sec:Width} is devoted to computation of the strong couplings $%
g_{Z_{q}B_{c}\rho}$ and $g_{Z_{s}B_{c}\phi }$. In this section we calculate
the widths of the decays $Z_{q}\rightarrow B_{c}\rho $ and $%
Z_{s}\rightarrow B_{c}\phi $. In Sec.\ \ref{sec:Disc} we examine
our results as a part of the general tetraquark's physics and compare them
with predictions of Ref. \cite{Chen:2013aba}, where the masses of the axial-vector
open charm-bottom tetraquarks were found. It contains also our concluding remarks.

\section{Masses and meson-current couplings}

\label{sec:Mass}

In order to find the masses and meson-current couplings of the
diquark-antidiquark type axial-vector states $Z_{q}$ and $Z_{s}$, we use the
two-point QCD sum rules. Below the explicit expressions for the $Z_{q}$
state are written down. Their generalization to embrace $Z_{s}$ \ tetraquark
is straightforward.

The two-point sum rule can be extracted from analysis of the correlation
function
\begin{equation}
\Pi _{\mu \nu }(p)=i\int d^{4}xe^{ipx}\langle 0|\mathcal{T}\{J_{\mu
}(x)J_{\nu }^{\dag }(0)\}|0\rangle ,  \label{eq:CorrF1}
\end{equation}%
where $J_{\mu }$ is the interpolating current of the $Z_{q}$ state.

The scalar and axial-vector open charm-bottom diquark-antidiquark states can
be modeled using different type of interpolating currents \cite{Chen:2013aba}.
Thus, the interpolating currents can be either symmetric or antisymmetric
in the color indices. In our previous work we chose the symmetric
interpolating current to find masses and decay widths of the scalar open
charm-bottom tetraquarks \cite{Agaev:2016dsg}. In the present work to
consider the axial-vector tetraquark states $Z_{q}$ and $Z_{s}$ we use again
the interpolating currents, which are symmetric in the color indices. Such
axial-vector current has the following form
\begin{equation}
J_{\mu }=q_{a}^{T}C\gamma _{5}c_{b}\left( \overline{q}_{a}\gamma _{\mu }C%
\overline{b}_{b}^{T}+\overline{q}_{b}\gamma _{\mu }C\overline{b}%
_{a}^{T}\right),  \label{eq:curr}
\end{equation}%
and is symmetric under exchange of the color indices $a\leftrightarrow b$.
Here by $C$ we denote the charge conjugation matrix.

To derive QCD sum rules for the mass and meson-current coupling we follow
standard prescriptions of the sum rule method and express the correlation
function $\Pi _{\mu \nu }(p)$ in terms of the physical parameters of the $%
Z_{q}$ state, which results in obtaining $\Pi _{\mu \nu }^{\mathrm{Phys}}(p)$%
. From another side the same function should be obtained in terms of the
quark-gluon degrees of freedom $\Pi _{\mu \nu }^{\mathrm{QCD}}(p)$.

We start from the function $\Pi _{\mu \nu }^{\mathrm{Phys}}(p)$ and compute
it by suggesting, that the tetraquarks under consideration are the ground
states in the relevant hadronic channels. After saturating the correlation
function with a complete set of the $Z_{q}$ states and performing in Eq.\ (%
\ref{eq:CorrF1}) integration over $x$ , we get the required expression for $%
\Pi _{\mu \nu }^{\mathrm{Phys}}(p)$
\begin{equation*}
\Pi _{\mu \nu }^{\mathrm{Phys}}(p)=\frac{\langle 0|J_{\mu }|Z_{q}(p)\rangle
\langle Z_{q}(p)|J_{\nu }^{\dag }|0\rangle }{m_{Z}^{2}-p^{2}}+...
\end{equation*}%
where $m_{Z}$ is the mass of the $Z_{q}$ state, and dots indicate
contributions coming from higher resonances and continuum states. We
introduce the meson-current coupling $f_{Z}$ by means of the equality%
\begin{equation*}
\langle 0|J_{\mu }|Z_{q}(p)\rangle =f_{Z}m_{Z}\varepsilon _{\mu },
\end{equation*}%
where $\varepsilon _{\mu }$ is polarization vector of the axial-vector
tetraquark. In terms of $m_{Z}$ and $f_{Z}$ the correlation function takes
the simple form
\begin{equation}
\Pi _{\mu \nu }^{\mathrm{Phys}}(p)=\frac{m_{Z}^{2}f_{Z}^{2}}{m_{Z}^{2}-p^{2}}%
\left( -g_{\mu \nu }+\frac{p_{\mu }p_{\nu }}{m_{Z}^{2}}\right) +\ldots
\label{eq:CorM}
\end{equation}%
Having applied the Borel transformation to the function $\Pi _{\mu \nu }^{%
\mathrm{Phys}}(p)$ we get
\begin{equation}
\mathcal{B}_{p^{2}}\Pi _{\mu \nu }^{\mathrm{Phys}%
}(p^{2})=m_{Z}^{2}f_{Z}^{2}e^{-m_{Z}^{2}/M^{2}}\left( -g_{\mu \nu }+\frac{%
p_{\mu }p_{\nu }}{m_{Z}^{2}}\right) +\ldots  \label{eq:CorBor}
\end{equation}

In order to obtain the function $\Pi _{\mu \nu }^{\mathrm{QCD}}(p)$ we
substitute the interpolating current given by Eq.\ (\ref{eq:curr}) into Eq.\
(\ref{eq:CorrF1}), and employ the light and heavy quark propagators in
calculations. For $\Pi_{\mu \nu }^{\mathrm{QCD}}(p)$, as a result, we get:
\begin{eqnarray}
&&\Pi _{\mu \nu }^{\mathrm{QCD}}(p)=i\int d^{4}xe^{ipx}\left\{ \mathrm{Tr}%
\left[ \gamma _{\mu }\widetilde{S}_{b}^{b^{\prime }b}(-x)\gamma _{\nu
}S_{q}^{a^{\prime }a}(-x)\right] \right.  \notag \\
&&\times \mathrm{Tr}\left[ \gamma _{5}\widetilde{S}_{q}^{aa^{\prime
}}(x)\gamma _{5}S_{c}^{bb^{\prime }}(x)\right] +\mathrm{Tr}\left[ \gamma
_{\mu }\widetilde{S}_{b}^{a^{\prime }b}(-x)\right.  \notag \\
&&\times \left. \gamma _{\nu }S_{q}^{b^{\prime }a}(-x)\right] \mathrm{Tr}%
\left[ \gamma _{5}\widetilde{S}_{q}^{aa^{\prime }}(x)\gamma
_{5}S_{c}^{bb^{\prime }}(x)\right]  \notag \\
&&+\mathrm{Tr}\left[ \gamma _{\mu }\widetilde{S}_{b}^{b^{\prime
}a}(-x)\gamma _{\nu }S_{q}^{a^{\prime }b}(-x)\right] \mathrm{Tr}\left[
\gamma _{5}\widetilde{S}_{q}^{aa^{\prime }}(x)\gamma _{5}S_{c}^{bb^{\prime
}}(x)\right]  \notag \\
&&\left. +\mathrm{Tr}\left[ \gamma _{\mu }\widetilde{S}_{b}^{a^{\prime
}a}(-x)\gamma _{\nu }S_{q}^{b^{\prime }b}(-x)\right] \mathrm{Tr}\left[
\gamma _{5}\widetilde{S}_{q}^{aa^{\prime }}(x)\gamma _{5}S_{c}^{bb^{\prime
}}(x)\right] \right\} ,  \notag \\
&&{}  \label{eq:CorrF2}
\end{eqnarray}%
where
\begin{equation}
\widetilde{S}_{q(b)}^{ab}(x)=CS_{q(b)}^{Tab}(x)C,  \label{eq:Not}
\end{equation}%
with $S_{q}(x)$ and $\ S_{b}(x)$ being the $q$- and $b$-quark propagators,
respectively.

We proceed including into analysis the well known expressions of the light
and heavy quark propagators. For our aims it is convenient to use the $x$%
-space expression of the light quark propagator,
\begin{eqnarray}
&&S_{q}^{ab}(x)=i\delta _{ab}\frac{\slashed x}{2\pi ^{2}x^{4}}-\delta _{ab}%
\frac{m_{q}}{4\pi ^{2}x^{2}}-\delta _{ab}\frac{\langle \overline{q}q\rangle
}{12}  \notag \\
&&+i\delta _{ab}\frac{\slashed xm_{q}\langle \overline{q}q\rangle }{48}%
-\delta _{ab}\frac{x^{2}}{192}\langle \overline{q}g_{s}\sigma Gq\rangle
+i\delta _{ab}\frac{x^{2}\slashed xm_{q}}{1152}\langle \overline{q}%
g_{s}\sigma Gq\rangle  \notag \\
&&-i\frac{g_{s}G_{ab}^{\alpha \beta }}{32\pi ^{2}x^{2}}\left[ \slashed x{%
\sigma _{\alpha \beta }+\sigma _{\alpha \beta }}\slashed x\right] -i\delta
_{ab}\frac{x^{2}\slashed xg_{s}^{2}\langle \overline{q}q\rangle ^{2}}{7776}
\notag \\
&&-\delta _{ab}\frac{x^{4}\langle \overline{q}q\rangle \langle
g_{s}q^{2}G^{2}\rangle }{27648}+\ldots . \label{eq:qprop}
\end{eqnarray}
For the heavy $Q=b,\ c$ quarks we utilize the propagator $S_{Q}^{ab}(x)$
given in the momentum space in Ref.\ \cite{Reinders:1984sr}:
\begin{eqnarray}
&&S_{Q}^{ab}(x)=i\int \frac{d^{4}k}{(2\pi )^{4}}e^{-ikx}\Bigg \{\frac{\delta
_{ab}\left( {\slashed k}+m_{Q}\right) }{k^{2}-m_{Q}^{2}}  \notag \\
&&-\frac{g_{s}G_{ab}^{\alpha \beta }}{4}\frac{\sigma _{\alpha \beta }\left( {%
\slashed k}+m_{Q}\right) +\left( {\slashed k}+m_{Q}\right) \sigma _{\alpha
\beta }}{(k^{2}-m_{Q}^{2})^{2}}  \notag \\
&&+\frac{g_{s}^{2}G^{2}}{12}\delta _{ab}m_{Q}\frac{k^{2}+m_{Q}{\slashed k}}{%
(k^{2}-m_{Q}^{2})^{4}}+\frac{g_{s}^{3}G^{3}}{48}\delta _{ab}\frac{\left( {%
\slashed k}+m_{Q}\right) }{(k^{2}-m_{Q}^{2})^{6}}  \notag \\
&&\times \left[ {\slashed k}\left( k^{2}-3m_{Q}^{2}\right) +2m_{Q}\left(
2k^{2}-m_{Q}^{2}\right) \right] \left( {\slashed k}+m_{Q}\right) +\ldots %
\Bigg \}.  \notag \\
&&{}  \label{eq:Qprop}
\end{eqnarray}%
In the expressions above
\begin{eqnarray}
&&G_{ab}^{\alpha \beta }=G_{A}^{\alpha \beta
}t_{ab}^{A},\,\,~~G^{2}=G_{\alpha \beta }^{A}G_{\alpha \beta }^{A},  \notag
\\
&&G^{3}=\,\,f^{ABC}G_{\mu \nu }^{A}G_{\nu \delta }^{B}G_{\delta \mu }^{C},
\end{eqnarray}%
where $a,\,b=1,2,3$ are  color indices and $A,B,C=1,\,2\,\ldots 8$. Here $%
t^{A}=\lambda ^{A}/2$ , where $\lambda ^{A}$ are the Gell-Mann matrices, and
the gluon field strength tensor is fixed at $x=0$, i.e. $G_{\alpha \beta
}^{A}\equiv G_{\alpha \beta }^{A}(0)$ $.$
\begin{table}[tbp]
\begin{tabular}{|c|c|}
\hline\hline
Parameters & Values \\ \hline\hline
$m_{B_c}$ & $(6275.1 \pm 1.0) ~\mathrm{MeV}$ \\
$f_{B_c}$ & $(528 \pm 19)~\mathrm{MeV}$ \\
$m_{\rho}$ & $(775.26 \pm 0.25) ~\mathrm{MeV}$ \\
$f_{\rho}$ & $216 \pm 3~\mathrm{MeV}$ \\
$m_{\phi}$ & $(1019.461 \pm 0.019) ~\mathrm{MeV}$ \\
$f_{\phi}$ & $215 \pm 5 ~\mathrm{MeV}$ \\
$m_{b}$ & $4.18^{+0.04}_{-0.03}~\mathrm{GeV}$ \\
$m_{c}$ & $(1.27 \pm 0.03)~\mathrm{GeV}$ \\
$m_{s} $ & $96^{+8}_{-4}~\mathrm{MeV} $ \\
$\langle \bar{q}q \rangle $ & $(-0.24\pm 0.01)^3$ $\mathrm{GeV}^3$ \\
$\langle \bar{s}s \rangle $ & $0.8\ \langle \bar{q}q \rangle$ \\
$m_{0}^2 $ & $(0.8\pm0.1)$ $\mathrm{GeV}^2$ \\
$\langle \overline{q}g_{s}\sigma Gq\rangle$ & $m_{0}^2\langle \bar{q}q
\rangle $ \\
$\langle \overline{s}g_{s}\sigma Gs\rangle$ & $m_{0}^2\langle \bar{s}s
\rangle $ \\
$\langle\frac{\alpha_sG^2}{\pi}\rangle $ & $(0.012\pm0.004)$ $~\mathrm{GeV}%
^4 $ \\
$\langle g_{s}^3G^3\rangle $ & $(0.57\pm0.29)$ $~\mathrm{GeV}^6 $ \\
\hline\hline
\end{tabular}%
\caption{Input parameters.}
\label{tab:Param}
\end{table}

The QCD sum rules can be derived after fixing the Lorentz structures in both
the physical and theoretical expressions of the correlation function and
equating the correspondent invariant functions. In the case of the
axial-vector particles the Lorentz structures in these expressions are ones $%
\sim g_{\mu \nu }$ and $\sim p_{\mu }p_{\nu }$. Because, the structures $%
\sim p_{\mu }p_{\nu }$ are contaminated by  the scalar
states with the same quark contents, we choose $\sim g_{\mu \nu }$ and the invariant
function $\Pi ^{\mathrm{QCD}}(p^{2})$ corresponding to this structure. Then in the theoretical
side of the sum rule there is only one invariant function $\Pi ^{\mathrm{QCD}%
}(p^{2})$, which can be represented as the dispersion integral
\begin{equation}
\Pi ^{\mathrm{QCD}}(p^{2})=\int_{\mathcal{M}^{2}}^{\infty }\frac{\rho ^{%
\mathrm{QCD}}(s)}{s-p^{2}}ds+...,  \label{eq:Sden}
\end{equation}%
where the lower limit of the integral $\mathcal{M}$ in the case under
consideration is equal to $\mathcal{M}=m_{b}+m_{c}$ . When considering the $%
Z_{s}$ state it should be replaced by $\mathcal{M}=m_{b}+m_{c}+2m_{s}$.

In Eq.\ (\ref{eq:Sden}), $\rho ^{\mathrm{QCD}}(s)$ is the spectral density
calculated as the imaginary part of the correlation function. It is the
important component of the sum rule calculations. Because the technical
tools necessary for derivation of $\rho ^{\mathrm{QCD}}(s)$ in the case of
the tetraquark states are well known and explained in the clear form in
Refs.\ \cite{Agaev:2016dev,Agaev:2016mjb}, here we avoid providing details
of relevant manipulations, and refrain also from presenting explicit
expressions for $\rho ^{\mathrm{QCD}}(s)$. We want to emphasize only that
the spectral density is computed by taking into account vacuum
condensates up to dimension eight, and include effects of the quark $\langle
\overline{q}q\rangle $, gluon  $\langle \alpha_{s}G^{2}/ \pi\rangle $, $\langle
g_{s}^{3}G^{3}\rangle $, mixed $\langle \overline{q}g_{s}\sigma Gq\rangle
$ condensates, and also terms of their products.

Applying the Borel transformation on the variable $p^{2}$ to the invariant
function $\Pi ^{\mathrm{QCD}}(p^{2})$, equating the obtained expression with
$\mathcal{B}_{p^{2}}\Pi ^{\mathrm{Phys}}(p)$, and subtracting the
contribution of higher resonances and continuum states, one finds the
required sum rule. Then the sum rule for the mass of the $Z_{q}$ state reads
\begin{equation}
m_{Z}^{2}=\frac{\int_{\mathcal{M}^{2}}^{s_{0}}ds\rho ^{\mathrm{QCD}%
}(s)se^{-s/M^{2}}}{\int_{\mathcal{M}^{2}}^{s_{0}}ds\rho ^{\mathrm{QCD}%
}(s)e^{-s/M^{2}}}.  \label{eq:srmass}
\end{equation}%
The meson-current coupling $f_{Z}$ can be extracted from the sum rule:
\begin{equation}
f_{Z}^{2}m_{Z}^{2}e^{-m_{Z}^{2}/M^{2}}=\int_{\mathcal{M}^{2}}^{s_{0}}ds\rho
^{\mathrm{QCD}}(s)e^{-s/M^{2}}.  \label{eq:srcoupling}
\end{equation}%
In Eqs.\ (\ref{eq:srmass}) and (\ref{eq:srcoupling}) by $s_{0}$ we denote
the threshold parameter, that separates the ground state's contribution from
contributions arising due to higher resonances and continuum.

\begin{widetext}

\begin{figure}[h!]
\begin{center}
\includegraphics[totalheight=6cm,width=8cm]{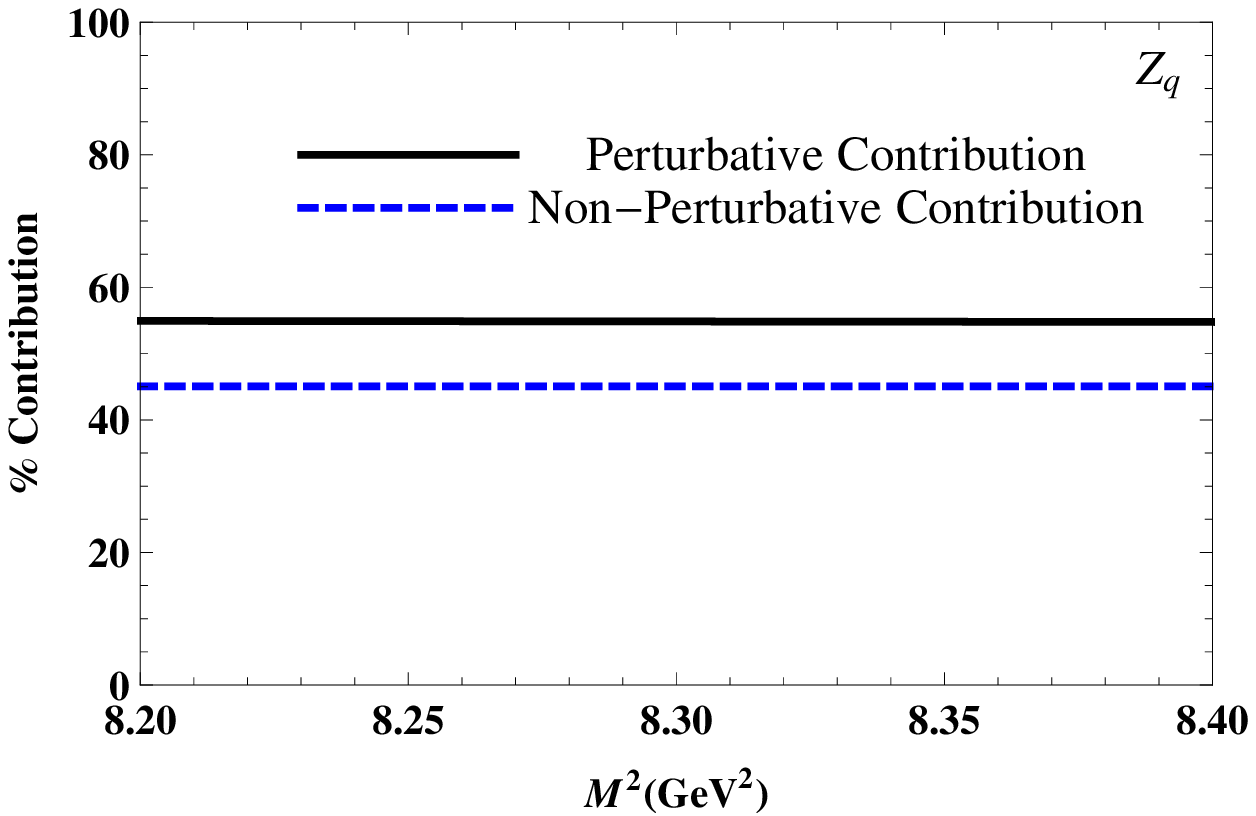}\,\,
\includegraphics[totalheight=6cm,width=8cm]{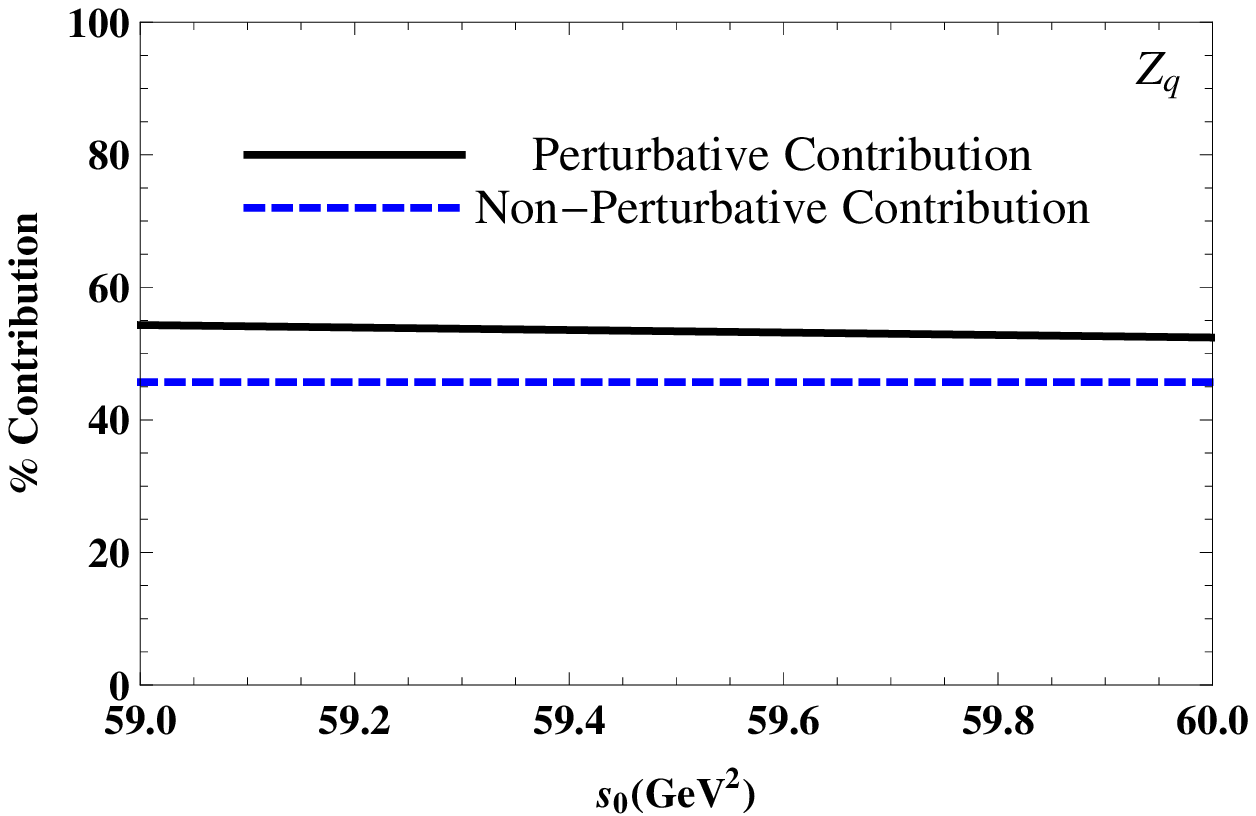}
\end{center}
\caption{  The perturbative and nonperturbative contributions to the sum rule
as functions of $M^2$ at an average $s_0$ (left panel), and as functions of $%
s_0$ at an average $M^2$ (right panel).}
\label{fig:Pert/Nonpert}
\end{figure}
\begin{figure}[h!]
\begin{center}
\includegraphics[totalheight=6cm,width=8cm]{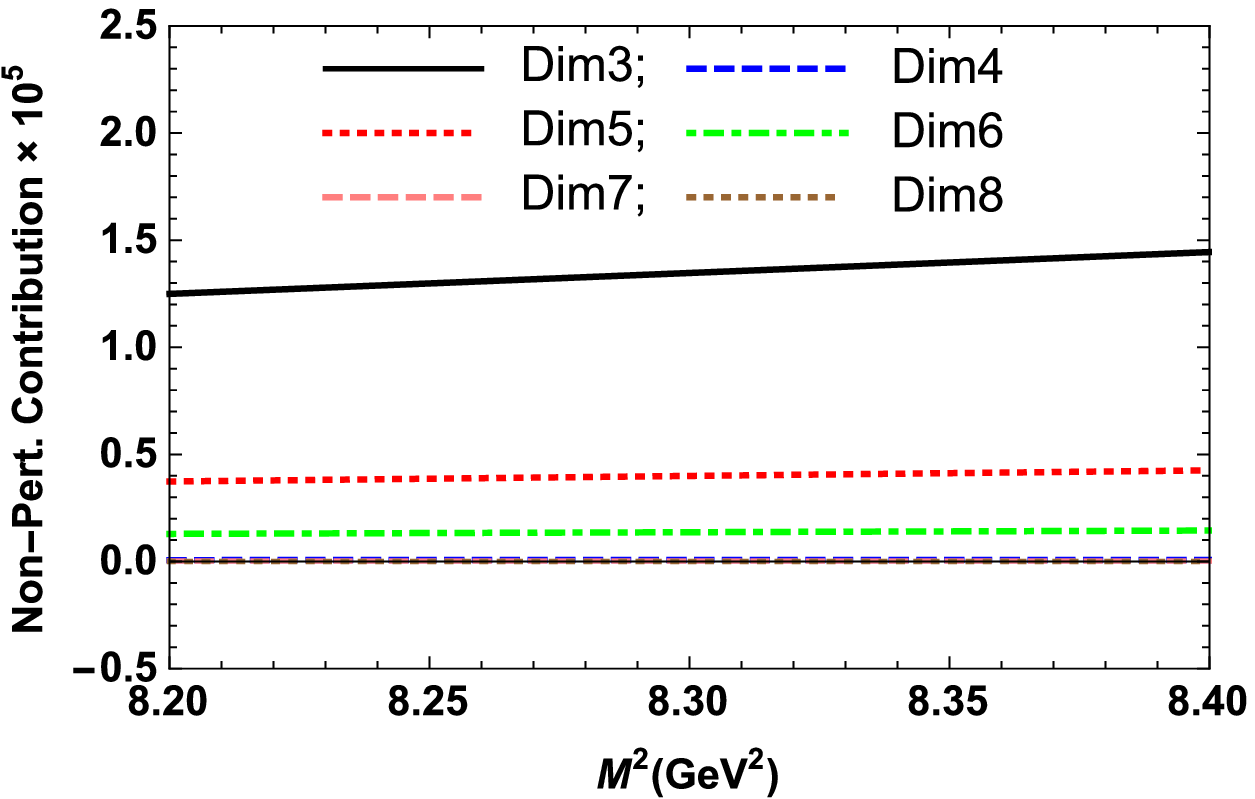}\,\, %
\includegraphics[totalheight=6cm,width=8cm]{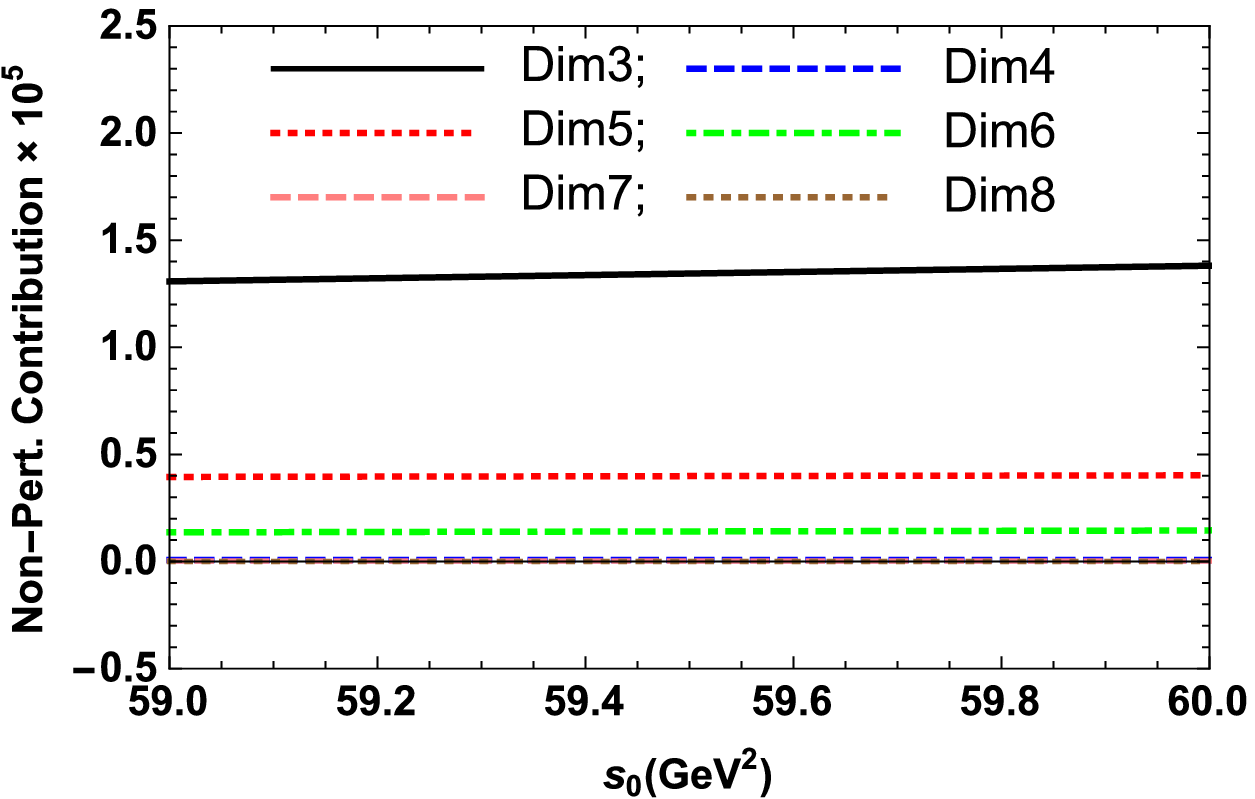}
\end{center}
\caption{  Contributions to the sum rule arising from the nonperturbative
operators of different dimensions are shown as functions of the Borel
parameter at an average value of $s_0$ (left panel), and as functions of the
threshold parameter $s_0$ at an average $M^2$ (right panel).}
\label{fig:Nonpert}
\end{figure}
\begin{figure}[h!]
\begin{center}
\includegraphics[totalheight=6cm,width=8cm]{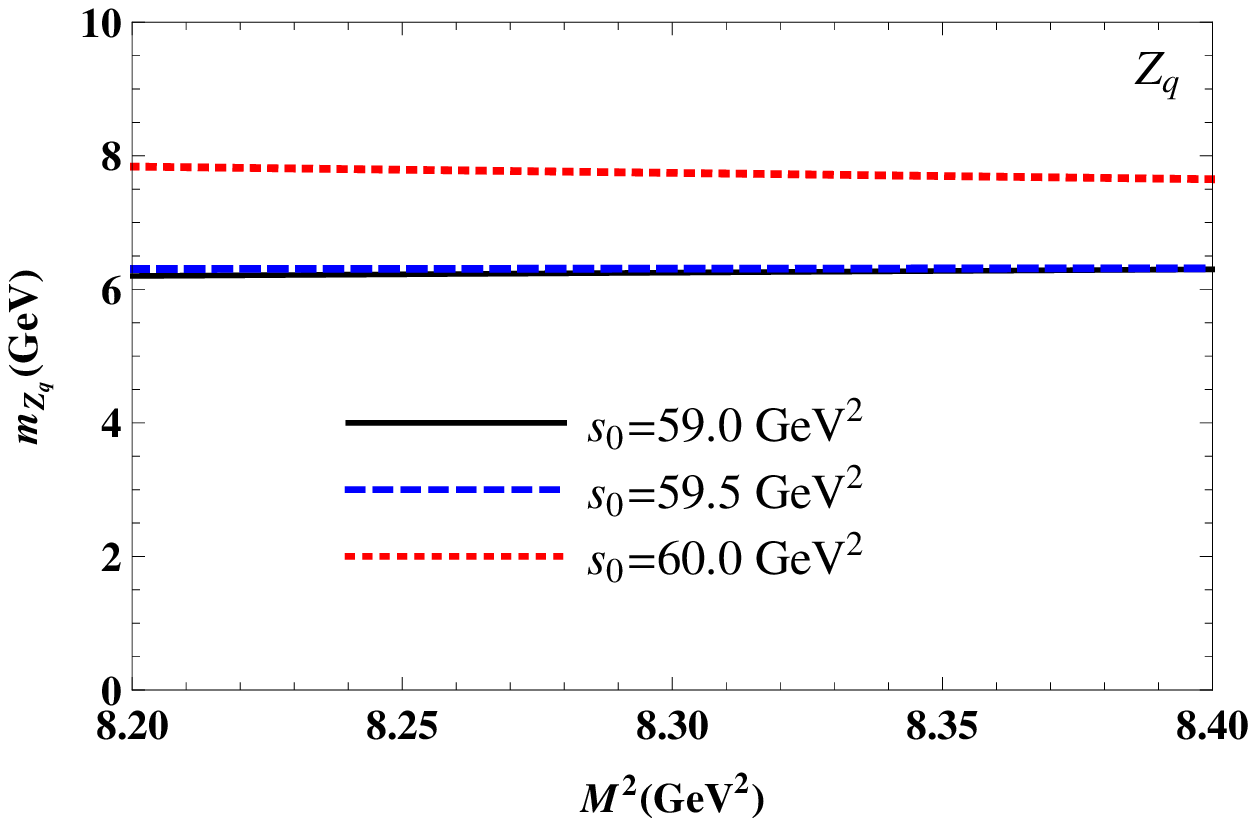}\,\, %
\includegraphics[totalheight=6cm,width=8cm]{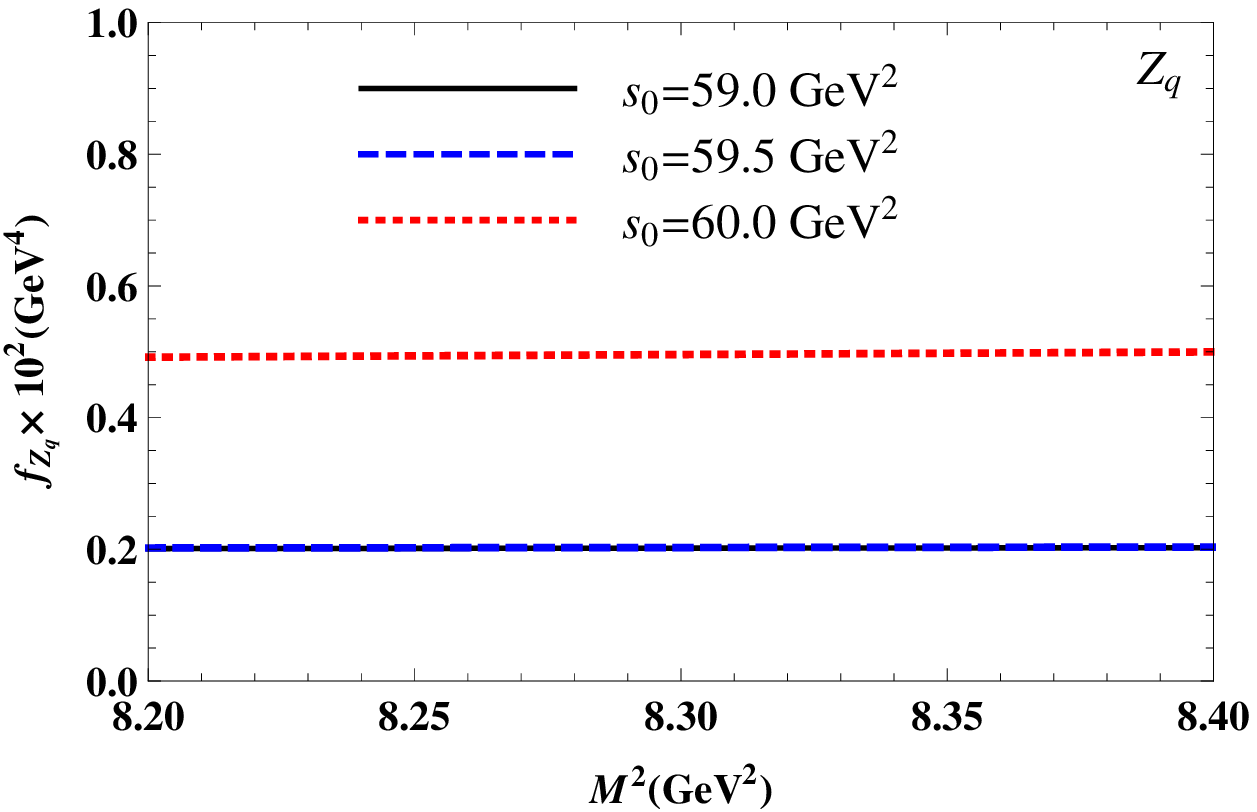}
\end{center}
\caption{ The mass (left panel) and meson-current coupling (right panel) as
functions of the Borel parameter $M^2$ at fixed values of the continuum
threshold $s_0$.}
\label{fig:MassCoup}
\end{figure}

\end{widetext}

The sum rules contain the parameters, which are necessary for numerical computations: Their
numerical values are collected in Table\ \ref{tab:Param}. The quark and gluon condensates are
well known, therefore we utilize their standard values. The Table\ \ref%
{tab:Param} contains also $B_{c}$, $\rho$ and $\phi $ mesons' masses (see,
Ref.\ \cite{Olive:2016xmw}) and decay constants, which will serve as input
parameters when computing the strong couplings and decay widths. It is worth
noting that for $f_{\rho}$, $\phi$ and $f_{B_c}$ we use the sum rule
estimations from Refs.\ \cite{Ball:2007zt,Baker:2013mwa}.

The sum rules Eqs.\ (\ref{eq:srmass}) and (\ref{eq:srcoupling}) contain also two
parameters $s_{0}$ and $M^{2}$, choices of which are decisive to extract
reliable estimations for the quantities under question. The continuum
threshold $s_0$ determines a boundary that dissects ground state contribution
from ones due to excited resonances and continuum. It depends on the energy
of the first excited state corresponding to the ground state hadron. The
continuum threshold $s_0$ can also be found from analysis of the pole to
total contribution ratio. The analysis done in the case of the tetraquark $%
Z_q$ allows us to fix a working interval for $s_0$ as
\begin{equation}
59\ \mathrm{GeV}^{2}\leq s_0\leq 60\ \mathrm{GeV}^{2}.  \label{eq:S0}
\end{equation}
The Borel parameter $M^2$ has also to satisfy well-known requirements.
Namely, convergence of OPE and exceeding of the perturbative contribution
over the nonperturbative one fixes a lower bound of the allowed values of $%
M^2$. The upper limit of the Borel parameter is determined to achieve the
largest possible pole contribution to the sum rule. These constraints lead
to the following working window for $M^2$
\begin{equation}
8.2\ \mathrm{GeV}^{2}\leq M^{2}\leq 8.4\ \mathrm{GeV}^{2}.
\end{equation}

In Figs.\ \ref{fig:Pert/Nonpert} and \ref{fig:Nonpert} we graphically
demonstrate some stages in extracting of the working regions for these
parameters. Thus, in Fig.\ \ref{fig:Pert/Nonpert} the perturbative and
nonperturbative contributions to the sum rule in the chosen regions for $s_0$
and $M^2$ are depicted. The convergence of OPE can be seen by inspecting
Fig.\ \ref{fig:Nonpert}, where the effects of the operators of the different
dimensions are plotted. By varying the parameters $s_0$ and $M^2$ within their
working ranges we find, that the pole contribution to the mass sum rule
amounts to $\sim 65\% $ of the result.

The final results for the mass and meson-current coupling of the $Z_q$ state
are drawn in Fig.\ \ref{fig:MassCoup} and collected in Table\ \ref%
{tab:Results1}. As is seen from Fig.\ \ref{fig:MassCoup}, the quantities
extracted from the sum rules demonstrate a mild dependence on $M^2$, whereas
effects of $s_0$ on them are sizable. The uncertainties generated by the
parameters $s_0$ and $M^2$ are main sources of errors, which are inherent
part of sum rule computations and equal up to $30\% $ of the whole integral.

The mass and meson-current coupling of the $Z_{s}$ state can be obtained
from the similar calculations, the difference being only in terms $\sim
m_{s} $ kept in the spectral density, whereas in $Z_{q}$ calculations we set
$m_{q}=0$. These modifications and also replacement $\mathcal{M}\Rightarrow
m_{b}+m_{c}+2m_{s}$ in the integrals result in shifting of the working
ranges of the parameters $s_0$ and $M^{2}$ towards slightly larger values,
which now read
\begin{eqnarray}
&&60\ \mathrm{GeV}^{2}\leq s_{0}\leq 61\ \mathrm{GeV}^{2},  \notag \\
&&8.4\ \mathrm{GeV}^{2}\leq M^{2}\leq 8.6\ \mathrm{GeV}^{2}.
\label{eq:Zspar}
\end{eqnarray}%
Predictions for $m_{Z_{s}}$ and $f_{Z_{s}}$ obtained using $s_0$ and $M^{2}$
from Eq.\ (\ref{eq:Zspar}) are also written down in Table\ \ref{tab:Results1}.

\begin{table}[tbp]
\begin{tabular}{|c|c|}
\hline\hline
Mass, m.-c. coupling & Results \\ \hline\hline
$m_{Z_q}$ & $(7.06\pm 0.74)~\mathrm{GeV}$ \\
$f_{Z_q} $ & $(0.33 \pm 0.11)\cdot 10^{-2}~\mathrm{GeV}^4 $ \\
$m_{Z_s}$ & $(7.30\pm 0.76)~\mathrm{GeV}$ \\
$f_{Z_s} $ & $(0.63 \pm 0.19)10^{-2}~\mathrm{GeV}^4 $ \\ \hline\hline
\end{tabular}%
\caption{The sum rule results for the masses and meson-current couplings of
the axial-vector $Z_{q}$ and $Z_s$ states.}
\label{tab:Results1}
\end{table}


\section{$Z_{q}\rightarrow B_{c}\protect\rho $ and $Z_{s}\rightarrow B_{c}%
\protect\phi $ decays}

\label{sec:Width}

In this section we investigate the strong decays of the exotic axial-vector $%
Z_{q(s)}$ states, and calculate widths of their main decay modes, which, in
accordance with results of Sec. \ref{sec:Mass}, are kinematically allowed.

One can see, that the quantum numbers, quark content and mass of the $Z_{q}$
tetraquark make the process $Z_{q}\rightarrow B_{c}\rho$ its preferable
decay mode. The $Z_{s}$ state may decay to $B_{c}$ and $\phi $ mesons. It is
worth noting that, due to the  $\rho -\omega $ and $\omega -\phi $
mixing, the processes $Z_{q}\rightarrow B_{c}\omega $ and $Z_{s}\rightarrow
B_{c}\omega $ are also among their kinematically allowed decay channels. But
because, for example, $\phi $ and $\omega $ mesons are almost pure $%
\overline{s}s$ and $\left( \overline{u}u+\overline{d}d\right) /\sqrt{2}$
states the $Z_{s}\rightarrow B_{c}\omega $ process is unessential provided
the mass of $Z_{s}$ allows its decay to $\phi $ meson: Alternative channels with
$\omega$ may play an important role in exploration of the tetraquark
states containing $\bar ss$ pair, if their masses are not enough to create
$\phi $ meson.

We are going to carry out a required analysis and write down all
expressions necessary to find the $Z_{q}\rightarrow B_{c}\rho $ decay's
width. After rather trivial replacements in corresponding formulas and input
parameters, the same calculations can easily be repeated for the $%
Z_{s}\rightarrow B_{c}\phi $ decay.

As first step we have to compute the coupling $g_{Z_{q}B_{c}\rho }$, which
describes the strong interaction in the vertex\ $Z_{q}B_{c}\rho $, and can
be extracted from the QCD sum rule. To this end, we
explore the correlation function
\begin{equation}
\Pi _{\mu }(p,q)=i\int d^{4}xe^{ipx}\langle \rho (q)|\mathcal{T}%
\{J^{B_{c}}(x)J_{\mu }^{\dag }(0)\}|0\rangle ,  \label{eq:CorrF3}
\end{equation}%
where $J^{B_{c}}(x)$ is the interpolating current of the $B_{c}$ meson: It
is defined in the form
\begin{equation}
J^{B_{c}}(x)=i\overline{b}_{l}(x)\gamma _{5}c_{l}(x).  \label{eq:Bcur}
\end{equation}%
The correlation function in Eq.\ (\ref{eq:CorrF3}) is introduced in the
form, which implies usage of the light-cone sum rule method. Indeed, $%
\Pi_{\mu}(p,q)$ will be computed employing QCD sum rule on the light-cone
by using a technique of the soft-meson approximation.

In terms of the physical parameters of the involved particles and coupling $%
g_{Z_{q}B_{c}\rho }$ the function $\Pi_{\mu}(p,q)$ has a simple form and
generates the phenomenological side of the sum rule. Namely,
\begin{eqnarray}
\Pi _{\mu }^{\mathrm{Phys}}(p,q) &=&\frac{\langle 0|J^{B_{c}}|B_{c}\left(
p\right) \rangle }{p^{2}-m_{B_{c}}^{2}}\langle B_{c}\left( p\right) \rho
(q)|Z_{q}(p^{\prime })\rangle  \notag \\
&&\times \frac{\langle Z_{q}(p^{\prime })|J_{\mu }^{\dagger }|0\rangle }{%
p^{\prime 2}-m_{Z}^{2}}+\ldots ,  \label{eq:CorrF4}
\end{eqnarray}%
where $p$, $q$ and $p^{\prime }=p+q$ are the momenta of $B_{c}$, $\rho $ and
$Z_{q}$ particles, respectively. The term presented above is the
contribution of the ground state: the dots stand for effects of the higher
resonances and continuum states.

We introduce the $B_{c}$ meson matrix element
\begin{equation*}
\langle 0|J^{B_{c}}|B_{c}\left( p\right) \rangle =\frac{%
f_{B_{c}}m_{B_{c}}^{2}}{m_{b}+m_{c}}
\end{equation*}%
where $m_{B_{c}}$ and $f_{B_{c}}$ are the mass and decay constant of the $%
B_{c}$ meson, and also the matrix element corresponding to the vertex
\begin{eqnarray}
\langle B_{c}\left( p\right) \rho (q)|Z_{q}(p^{\prime })\rangle
&=&g_{Z_{q}B_{c}\rho }\left[ \left( q\cdot \varepsilon ^{\prime }\right)
\left( p^{\prime }\cdot \varepsilon ^{\ast }\right) \right.  \notag \\
&&\left. -\left( q\cdot p^{\prime }\right) \left( \varepsilon ^{\ast }\cdot
\varepsilon ^{\prime }\right) \right] .  \label{eq:Mel}
\end{eqnarray}%
Then the ground state term in the correlation function can be easily found,
as:
\begin{eqnarray}
&&\Pi _{\mu }^{\mathrm{Phys}}(p,q)=\frac{%
f_{B_{c}}f_{Z}m_{Z}m_{B_{c}}^{2}g_{Z_{q}B_{c}\rho }}{\left( p^{\prime
2}-m_{Z}^{2}\right) \left( p^{2}-m_{B_{c}}^{2}\right) (m_{b}+m_{c})}  \notag
\\
&&\times \left( \frac{m_{Z}^{2}-m_{B_{c}}^{2}}{2}\varepsilon _{\mu }^{\ast
}-p^{\prime }\cdot \varepsilon ^{\ast }q_{\mu }\right) +\ldots .
\label{eq:CorrF5}
\end{eqnarray}%

Strong vertices of a tetraquark with two conventional mesons differ from vertices
containing only ordinary mesons. The reason here is very simple: the tetraquark
$Z_q$  is a state composed of four
valence quarks, therefore the expansion of the non-local
correlation function $\Pi _{\mu }(p,q)$ leads to the expression, which instead of distribution
amplitudes of  $\rho$ meson depends on its local matrix elements (of course,  same arguments are valid for $Z_s$, as well). Then, the conservation of
the four-momentum at the vertex $Z_qB_c\rho$ equals $q$ to zero. In other words, within the
light-cone sum rule method the momentum of  $\rho$ meson should be equal to zero in our case.
In  vertices of ordinary hadrons four-momenta  of all involved particles can take nonzero values.
The soft-meson approximation corresponds to a situation when $q=0$. Calculations
of the same strong couplings within the full light-cone sum rule method and in the soft-meson
approximation demonstrated that the difference between results extracted using these two approaches
is numerically small (for detailed discussion, see Ref.\ \cite{Braun:1995}).

In the soft limit $p=p^{\prime }$,  only the term that survives
in Eq.\ (\ref{eq:CorrF5}) is $\sim \varepsilon _{\mu }^{\ast }$.  The invariant function $\Pi ^{\mathrm{Phys}}(p^{2})$ corresponding to this structure depends on the variable $p^{2}$, and is given as
\begin{eqnarray}
&&\Pi ^{\mathrm{Phys}}(p^{2})=\frac{%
f_{B_{c}}f_{Z}m_{Z}m_{B_{c}}^{2}g_{Z_{s}B_{c}\eta }}{2\left(
p^{2}-m^{2}\right) ^{2}(m_{b}+m_{c})}  \notag \\
&&\times \left( m_{Z}^{2}-m_{B_{c}}^{2}\right) +\ldots ,  \label{eq:CorrF5A}
\end{eqnarray}%
where $m^{2}=\left( m_{Z}^{2}+m_{B_{c}}^{2}\right) /2.$

In the soft-meson approximation we additionally apply the operator
\begin{equation}
\left( 1-M^{2}\frac{d}{dM^{2}}\right) M^{2}e^{m^{2}/M^{2}},
\label{eq:softop}
\end{equation}%
to both sides of the sum rule. The last operation is required to remove all
unsuppressed contributions existing in the physical side of the sum rule in
the soft-meson limit (see, Ref.\ \cite{Ioffe:1983ju}).

The second component of the sum rule, i.e. QCD expression for the
correlation function $\Pi^{\mathrm{QCD}}_{\mu}(p,q)$ is calculated employing
the quark propagators and shown below
\begin{eqnarray}
&&\Pi _{\mu }^{\mathrm{QCD}}(p,q)=-i\int d^{4}xe^{ipx}\left\{ \left[ \gamma
_{5}\widetilde{S}_{c}^{ib}(x){}\gamma _{5}\right. \right.  \notag \\
&&\left. \times \widetilde{S}_{b}^{bi}(-x){}\gamma _{\mu }\right] _{\alpha
\beta }\langle \rho (q)|\overline{q}_{\alpha }^{a}q_{\beta }^{a}|0\rangle
\notag \\
&&\left. +\left[ \gamma _{5}\widetilde{S}_{c}^{ib}(x)\gamma _{5}\widetilde{S}%
_{b}^{ai}(-x){}\gamma _{\mu }\right] _{\alpha \beta }\langle \rho (q)|%
\overline{s}_{\alpha }^{a}s_{\beta }^{b}|0\rangle \right\} ,
\end{eqnarray}%
with $\alpha $ and $\beta $ being the spinor indices.

We continue our calculations by employing the expansion
\begin{equation}
\overline{q}_{\alpha }^{a}q_{\beta }^{b}\rightarrow \frac{1}{4}\Gamma
_{\beta \alpha }^{j}\left( \overline{q}^{a}\Gamma ^{j}q^{b}\right) ,
\label{eq:MatEx}
\end{equation}%
where $\Gamma ^{j}=1,\ \gamma _{5},\ \gamma _{\mu },\ i\gamma _{5}\gamma
_{\mu },\ \sigma _{\mu \nu }/\sqrt{2}$ is the full set of Dirac matrices,
and carry out the color summation.

Prescriptions to perform summation over color indices, as well as procedures
to calculate resulting integrals and extract the imaginary part of the
correlation function $\Pi _{\mu }^{\mathrm{QCD}}(p,q)$ were numerously
presented in our previous works Refs.\ \cite%
{Agaev:2016dev,Agaev:2016ijz,Agaev:2016lkl,Agaev:2016urs}. Therefore, here
we skip further details, and provide the $\rho $ meson local matrix elements
that in the soft limit contribute to the spectral density, as well as,
final formulas for the spectral density $\rho _{c}(s)$.

Analysis demonstrates that in the soft limit only the matrix elements
\begin{equation}
\langle 0|\overline{q}\gamma _{\mu }q|\rho ^{0}(p)\rangle =\frac{1}{\sqrt{2}}%
f_{\rho }m_{\rho }\varepsilon _{\mu },  \label{eq:MatE2}
\end{equation}%
and%
\begin{equation}
\langle 0|\overline{q}g\widetilde{G}_{\mu \nu }\gamma _{\nu }\gamma
_{5}q|\rho ^{0}(p)\rangle =\frac{1}{\sqrt{2}}f_{\rho }m_{\rho }^{3}\zeta
_{4\rho }\varepsilon _{\mu },  \label{eq:TW4}
\end{equation}
are involved into computations, where $q$ denotes one of the $u$ or $d$
quarks. The matrix elements depend on the $\rho $ meson mass $m_{\rho }$ and
decay constant $f_{\rho }$. The twist-4 matrix element in Eq.\ (\ref{eq:TW4}%
), as a factor, contains also the parameter $\zeta _{4\rho }$. Its numerical
value was extracted at the scale $\mu =1\ \mathrm{GeV}$ from the sum rule
calculations in Ref.\ \cite{Ball:2007zt} and equals to%
\begin{equation*}
\zeta _{4\rho }=0.07\pm 0.03.
\end{equation*}

The final expression of the spectral density has the form
\begin{equation}
\rho _{c}(s)=\frac{f_{\rho }m_{\rho }}{24\sqrt{2}}\left[ F^{\mathrm{pert.}%
}(s)+F^{\mathrm{n.-pert.}}(s)\right] .
\end{equation}%
Here $F^{\mathrm{pert.}}(s)$ is the perturbative contribution to $\rho
_{c}(s)$
\begin{eqnarray}
&&F^{\mathrm{pert.}}(s) =\frac{1}{\pi ^{2}s^{2}}\left\{ \left[ s^{2}+s\left(
m_{b}^{2}+6m_{b}m_{c}+m_{c}^{2}\right) \right. \right.  \notag \\
&&\left. \left. -2(m_{b}^{2}-m_{c}^{2})^{2}\right] \right\} \sqrt{\left(
s+m_{b}^{2}-m_{c}^{2}\right) ^{2}-4m_{b}^{2}s},
\end{eqnarray}%
whereas by $F^{\mathrm{n.-pert.}}(s)$ we denote its nonperturbative
component. The function $F^{\mathrm{n.-pert.}}(s)$ is the sum of the terms
\begin{eqnarray}
&&F^{\mathrm{n.-pert.}}(s)=F_{G}^{\mathrm{n.-pert.}}(s)+\Big \langle\frac{%
\alpha _{s}G^{2}}{\pi }\Big \rangle\int_{0}^{1}f_{g_{s}^{2}G^{2}}(z,s)dz
\notag \\
&&+\Big \langle g_{s}^{3}G^{3}\Big \rangle%
\int_{0}^{1}f_{g_{s}^{3}G^{3}}(z,s)dz  \notag \\
&&+\Big \langle\frac{\alpha _{s}G^{2}}{\pi }\Big \rangle^{2}%
\int_{0}^{1}f_{(g_{s}^{2}G^{2})^{2}}(z,s)dz.  \label{eq:NPert}
\end{eqnarray}
Here $F_{G}^{\mathrm{n.-pert.}}(s)$ appears from integration of the
perturbative component of one heavy quark propagator with the term $\sim G$
from another one. It can be expressed using the matrix element given by Eq.\
(\ref{eq:TW4}) and has a rather simple form
\begin{equation}
F_{G}^{\mathrm{n.-pert.}}(s)=\frac{3m_{\rho }^{2}\zeta _{4\rho }}{2\pi ^{2}s}%
\sqrt{\left( s+m_{b}^{2}-m_{c}^{2}\right) ^{2}-4m_{b}^{2}s}.
\end{equation}%
The nonperturbative factors in front of the integrals, and subscripts of the functions
clearly indicate the origin of the remaining terms. In fact, the functions $f_{g_{s}^{2}G^{2}}$, $%
f_{g_{s}^{3}G^{3}}$ are due to products of $\sim g_{s}^{2}G^{2}$ and $\sim
g_{s}^{3}G^{3}$ terms with the perturbative component of another propagator,
whereas $f_{(g_{s}^{2}G^{2})^{2}}$ comes from integrals obtained using $\sim
g_{s}^{2}G^{2}$ components of $b$ and $c$ quarks' propagators. These terms
are four, six and eight dimensional nonperturbative contributions to the
spectral density $\rho _{c}(s)$, respectively. Their explicit forms are
presented below:
\begin{widetext}
\begin{eqnarray}
&&f_{g_{s}^{2}G^{2}}(z,s)=\frac{1}{12z^{2}(z-1)^{2}}\left\{
54(1-z)z^{2}\delta (s-\Phi )+\left[ 8m_{b}^{2}(z-1)^{3}+z^{2}\left(
27s(1-z)-8m_{b}^{2}z\right) \right. \right.   \notag \\
&&\left. \left. +2m_{b}m_{c}\left( 4+15z+12z^{2}\right) \right] \delta
^{(1)}(s-\Phi )-4s\left[ m_{b}^{2}(1-z)^{3}+m_{b}m_{c}z(1-z)-m_{c}^{2}z^{3}%
\right] \delta ^{(2)}(s-\Phi )\right\} ,
\end{eqnarray}%
\begin{eqnarray}
&&f_{g_{s}^{3}G^{3}}(z,s)=\frac{1}{15\cdot 2^{6}z^{5}(z-1)^{5}}\left\{
-12z^{2}(z-1)^{2}\left[ 3m_{b}^{2}(z-1)^{5}+3m_{b}m_{c}((1-z)^{5}+z^{5})+z%
\left( -3m_{c}^{2}z^{4}\right. \right. \right.   \notag \\
&&\left. \left. +s\left( 1-8z+25z^{2}-40z^{3}+33z^{4}-11z^{5}\right) \right)
\right] \delta ^{(2)}(s-\Phi )+2z(z-1)\left[ m_{b}^{2}(z-1)^{5}\left(
7m_{b}^{2}-4m_{b}m_{c}-9sz(2z-1)\right) \right.   \notag \\
&&\left. +2m_{b}m_{c}z^{2}\left(
2m_{c}^{2}z^{3}-9s(z-1)^{2}(1-3z+3z^{2})\right) +z^{3}\left(
-7m_{c}^{4}z^{2}+9sm_{c}^{2}z^{2}(1-3z+2z^{2})+2s^{2}(z-1)^{3}(2-7z+7z^{2})%
\right) \right]   \notag \\
&&\times \delta ^{(3)}(s-\Phi )+\left[
-2m_{b}^{5}m_{c}(z-1)^{5}+7m_{b}^{4}sz(z-1)^{6}-4m_{b}^{3}m_{c}sz^{2}(z-1)^{5}
-6m_{b}^{2}s^{2}z^{3}(z-1)^{6}+2m_{b}m_{c}z^{4}\right.
\notag \\
&&\left. \left. \times \left(
-3s^{2}(z-1)^{4}+m_{c}^{4}z+2m_{c}^{2}sz(z-1)^{2}\right) +s(z-1)z^{5}\left(
s^{2}(z-1)^{4}-7m_{c}^{4}z+6m_{c}^{2}sz(z-1)^{2}\right) \right] \delta
^{(4)}(s-\Phi )\right\} ,
\end{eqnarray}%
\begin{equation}
f_{(g_{s}^{2}G^{2})^{2}}(z,s)=\frac{m_{b}m_{c}}{54z^{2}(z-1)^{2}}\left\{ 2%
\left[ m_{b}m_{c}-s(1-3z+3z^{2})\right] \delta ^{(4)}(s-\Phi
)+s[m_{b}m_{c}+s(1-z)z]\delta ^{(5)}(s-\Phi )\right\} ,
\end{equation}
\end{widetext}
where,
\begin{equation*}
\delta^{(n)}(s-\Phi )=\frac{d^{n}}{ds^{n}}\delta(s-\Phi ),
\end{equation*}
with $\Phi $ being defined as
\begin{equation*}
\Phi =\frac{m_{b}^{2}(1-z)+m_{c}^{2}z}{z(1-z)}.
\end{equation*}

The final sum rule to evaluate the strong coupling reads
\begin{eqnarray}
&&g_{Z_{q}B_{c}\rho }=\frac{2(m_{b}+m_{c})}{%
f_{B_{c}}f_{Z}m_{Z}m_{B_{c}}^{2}(m_{Z}^{2}-m_{B_{c}}^{2})} \left( 1-M^{2}%
\frac{d}{dM^{2}}\right)  \notag \\
&&\times M^{2}\int_{(m_b+m_{c})^2}^{s_{0}}dse^{(m^{2}-s)/M^{2}}\rho
_{c}(s).  \label{eq:SRules}
\end{eqnarray}%
To calculate the width of the decay $Z_{q}\rightarrow B_{c}\rho $ we use the
expression,
\begin{eqnarray}
&&\Gamma \left( Z_{q}\rightarrow B_{c}\rho \right) =\frac{g_{Z_{q}B_{c}\rho
}^{2}m_{\rho}^{2}}{24\pi }\lambda \left( m_{Z},\ m_{B_{c}},m_{\rho }\right)
\notag \\
&&\times \left[ 3+\frac{2\lambda ^{2}\left( m_{Z_{q}},\ m_{B_{c}},m_{\rho
}\right) }{m_{\rho}^{2}}\right] ,  \label{eq:DW}
\end{eqnarray}%
where
\begin{equation*}
\lambda (a,\ b,\ c)=\frac{\sqrt{a^{4}+b^{4}+c^{4}-2\left(
a^{2}b^{2}+a^{2}c^{2}+b^{2}c^{2}\right) }}{2a}.
\end{equation*}

Parameters necessary for numerical calculations of the strong coupling $%
g_{Z_{q}B_{c}\rho }$ and $\Gamma \left( Z_{q}\rightarrow B_{c}\rho \right) $
are listed in Table\ \ref{tab:Param}.
\begin{table}[tbp]
\begin{tabular}{|c|c|}
\hline\hline
Strong couplings, Widths & Predictions \\ \hline\hline
$g_{Z_{q}B_c \rho}$ & $(5.31 \pm 1.25) $ $\mathrm{GeV}^{-1}$ \\
$g_{Z_{s}B_c \phi}$ & $(6.42 \pm 1.52) $ $\mathrm{GeV}^{-1}$ \\
$\Gamma(Z_q \to B_c \rho)$ & $(80 \pm 32)$ $\mathrm{MeV}$ \\
$\Gamma(Z_s \to B_c \phi)$ & $(168 \pm 68)$ $\mathrm{MeV}$ \\ \hline\hline
\end{tabular}%
\caption{The strong couplings and decay widths of the $Z_{q}$ and $Z_{s}$
tetraquarks. }
\label{tab:Results2}
\end{table}
The investigation carried out in accordance with standard requirements of
the sum rule calculations allows us to determine the ranges for $s_{0}$ and $%
M^{2}$. For example, the pole contribution to the sum rule amounts to $\sim
48-60 \%$ of the total result, as is seen from Fig.\ \ref{fig:PCcoupling}.
Other constraints, i.e. convergence of OPE, prevalence of the perturbative
contribution have been checked, as well. Summing up the performed analysis
we fix the interval for the continuum threshold $s_{0}$ as in the mass
calculations (see, Eq.\ (\ref{eq:S0})), whereas  for the
Borel parameter we obtain
\begin{equation}
8\ \mathrm{GeV}^{2}\leq M^{2}\leq 9\ \mathrm{GeV}^{2},
\end{equation}
which is wider than the corresponding window in the mass sum rule.
\begin{widetext}

\begin{figure}[h!]
\begin{center}
\includegraphics[totalheight=6cm,width=8cm]{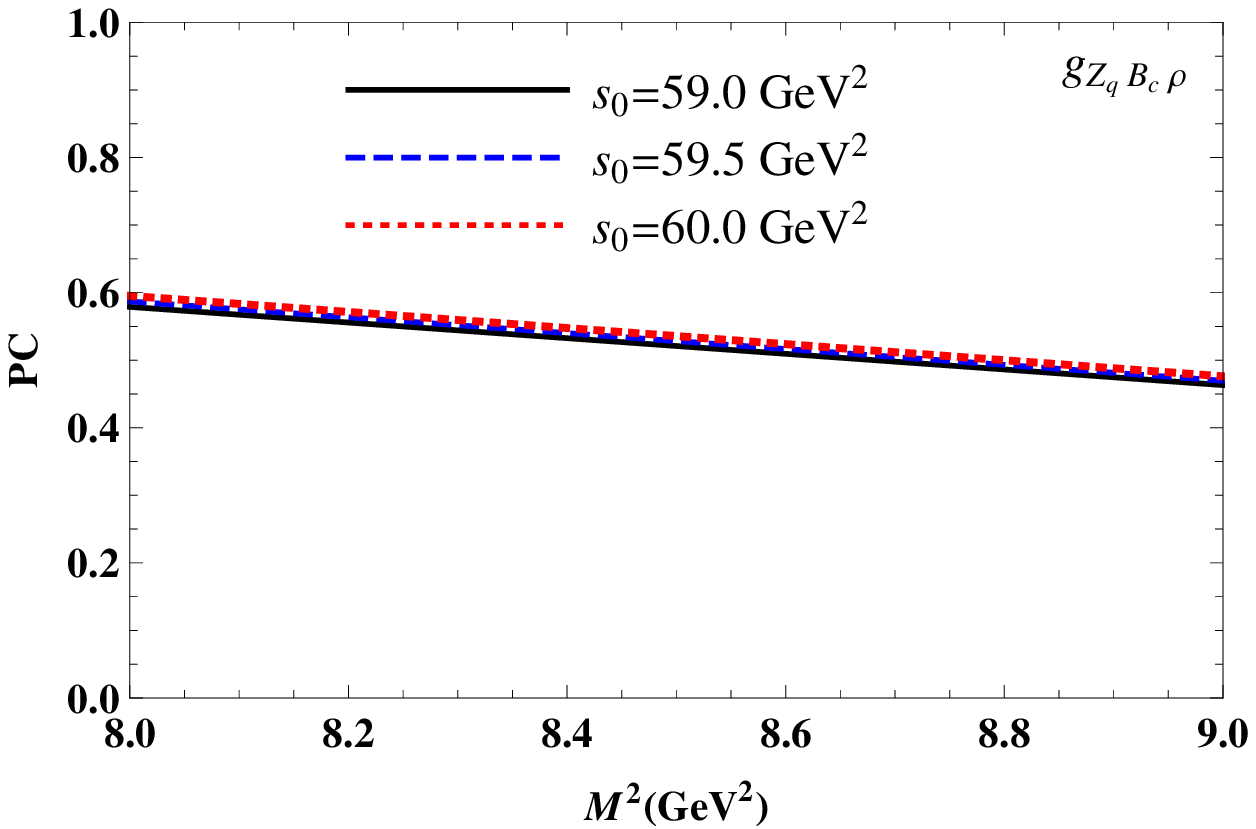}\,\, %
\includegraphics[totalheight=6cm,width=8cm]{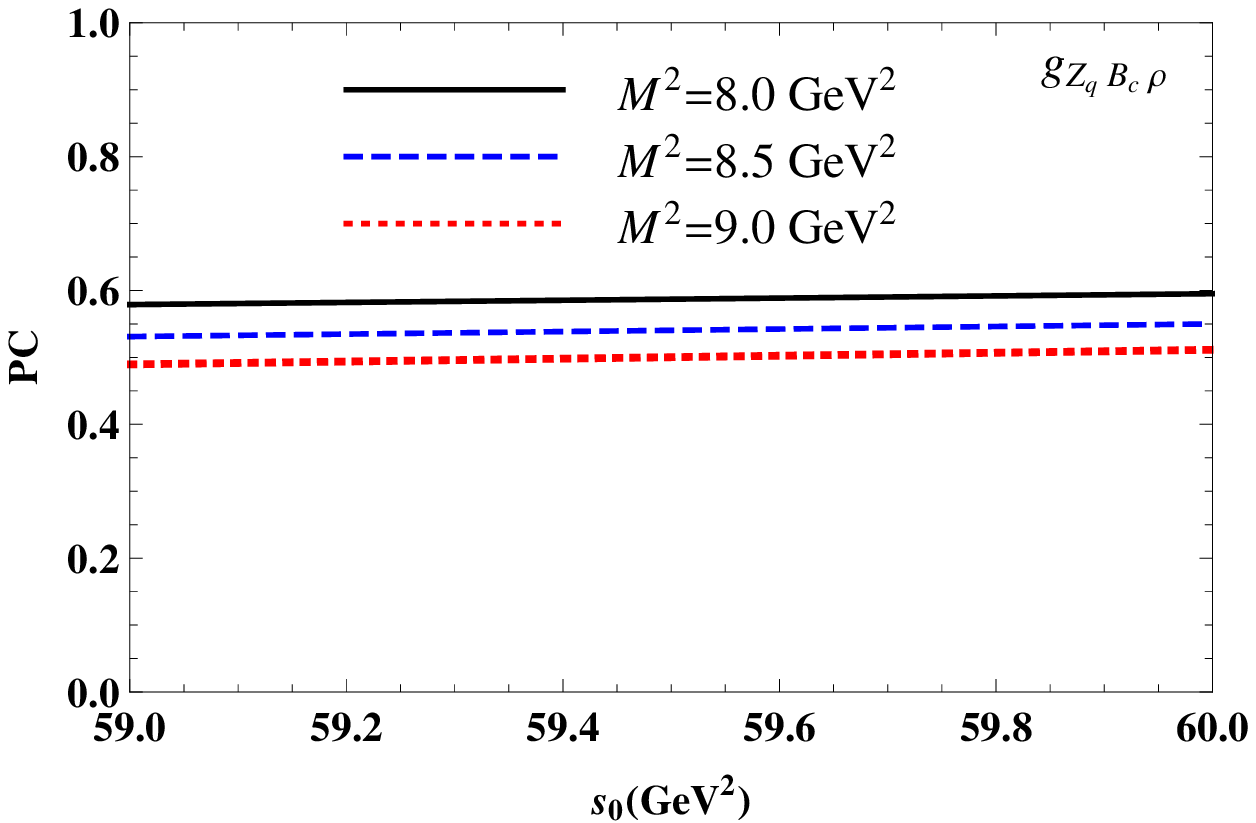}
\end{center}
\caption{ The pole contribution in the $g_{Z_qB_c\protect\rho}$ coupling sum
rule calculations as a function of the Borel parameter $M^2$ at fixed $s_0$
(left panel), and as a function of the threshold $s_0$ at fixed values of $%
M^2 $ (right panel).}
\label{fig:PCcoupling}
\end{figure}
\begin{figure}[h!]
\begin{center}
\includegraphics[totalheight=6cm,width=8cm]{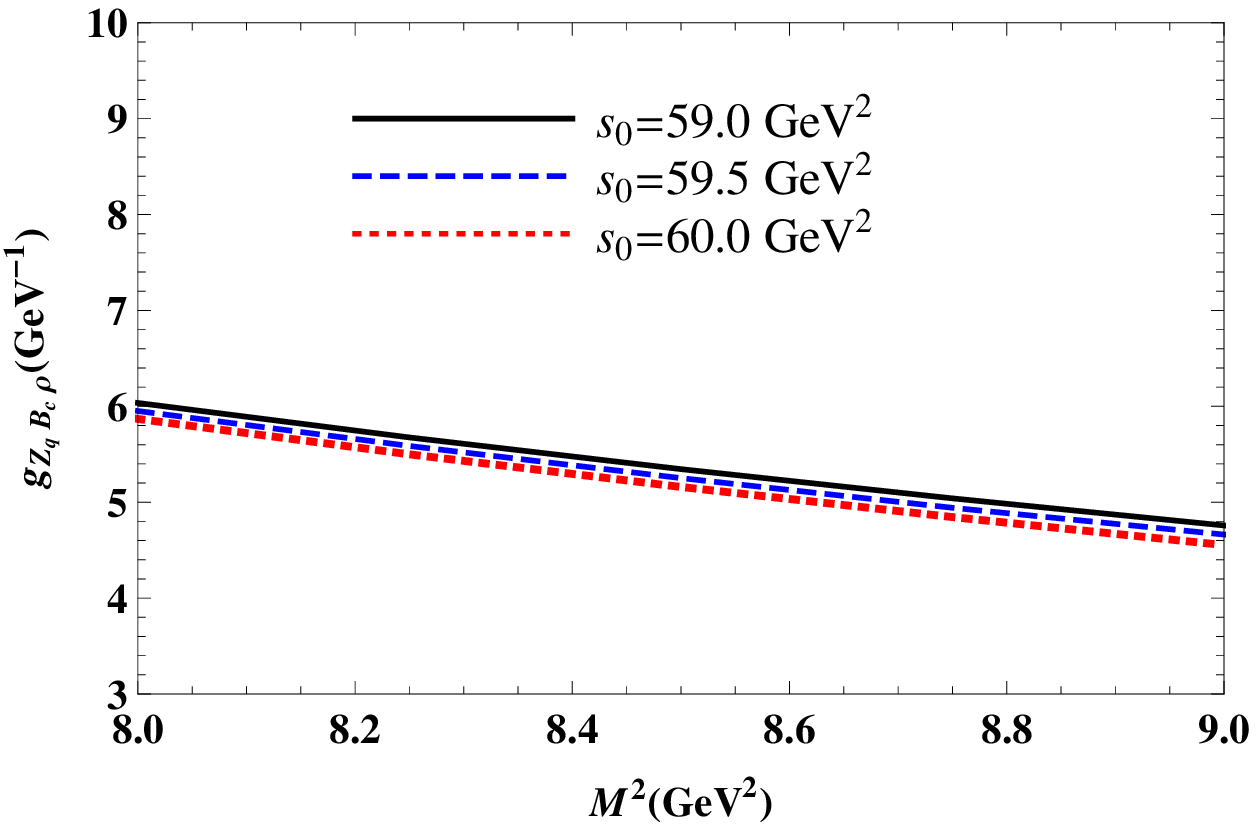}\,\, %
\includegraphics[totalheight=6cm,width=8cm]{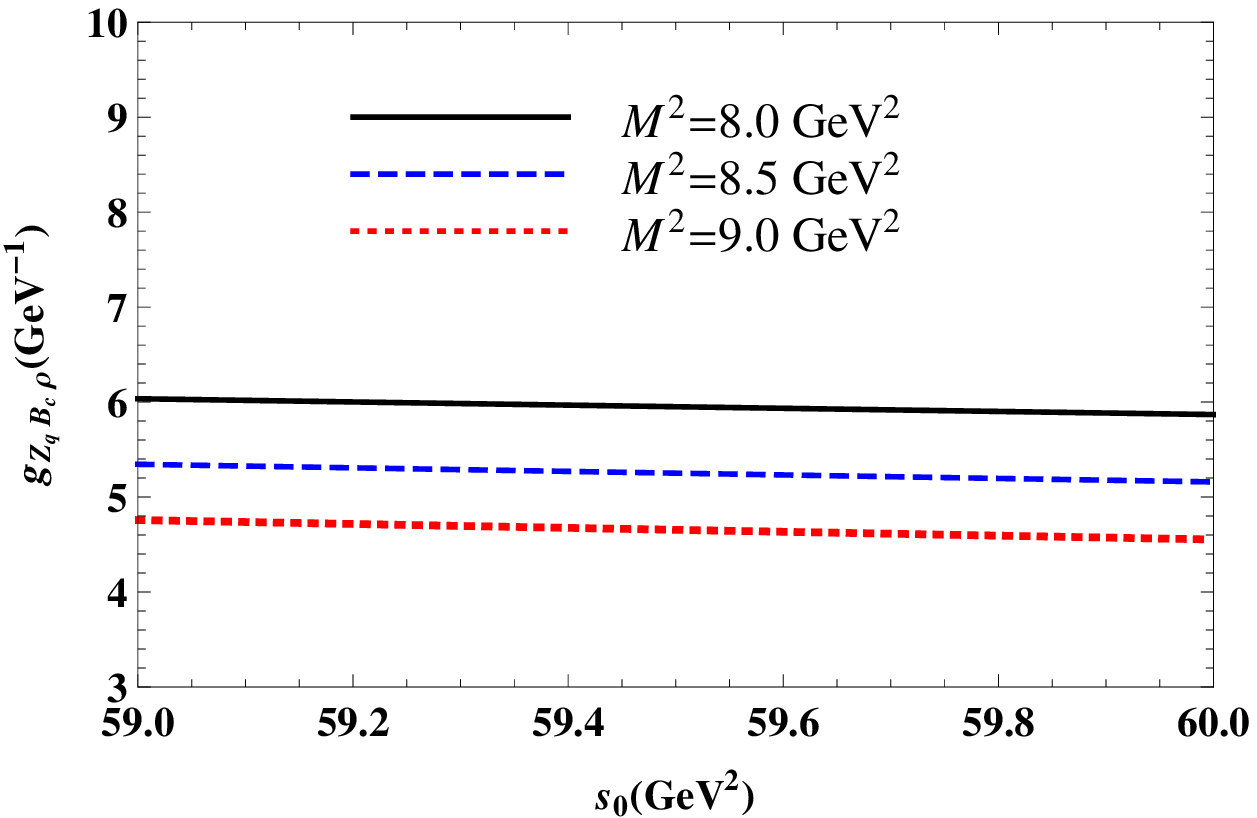}
\end{center}
\caption{ The strong coupling $g_{Z_qB_c\protect\rho}$ as a function of the
Borel parameter $M^2$ at fixed $s_0$ (left panel), and as a function of the
threshold $s_0$ at fixed values of $M^2$ (right panel).}
\label{fig:Coupling}
\end{figure}

\end{widetext}

In Fig.\ \ref{fig:Coupling} we provide our final results and depict the
strong coupling $g_{Z_qB_c\rho}$ as the function of the Borel parameter (at
fixed $s_{0}$) and as the function of the continuum threshold (at fixed $M^2$).
The dependence of the strong coupling on these parameters has a traditional form,
and systematic errors of the calculations are within reasonable limits.

The decay $Z_{s}\rightarrow B_{c}\phi $ can be considered in analogous
manner: One only needs  to write down in the relevant expressions the parameters of the
$\phi$ meson. Thus, the matrix elements of the $\phi $ meson that take
part in forming of the spectral density are
\begin{eqnarray*}
\langle 0|\overline{s}\gamma _{\mu }s|\phi (p)\rangle  &=&f_{\phi }m_{\phi
}\varepsilon _{\mu }, \\
\langle 0|\overline{s}g\widetilde{G}_{\mu \nu }\gamma _{\nu }\gamma
_{5}s|\phi (p)\rangle  &=&f_{\phi }m_{\phi }^{3}\zeta _{4\phi }\varepsilon
_{\mu },
\end{eqnarray*}
where the twist-4 parameter%
\begin{equation*}
\zeta _{4\phi }=0.00\pm 0.02
\end{equation*}%
was estimated and found compatible with zero in  Ref.\ \cite{Ball:2007zt}.

In calculations of the coupling $g_{Z_{s}B_{c}\phi }$ the working regions for
the Borel parameter and continuum threshold   are fixed in the form:
\begin{eqnarray}
&&60\ \mathrm{GeV}^{2}\leq s_{0}\leq 61\ \mathrm{GeV}^{2},  \notag \\
&&8.2\ \mathrm{GeV}^{2}\leq M^{2}\leq 9.2\ \mathrm{GeV}^{2}.
\end{eqnarray}%
Our results for the strong couplings and
widths of the  decay modes studied in this work are collected in Table\ \ref%
{tab:Results2}.


\section{Discussion and concluding remarks}

\label{sec:Disc}
In the present work we have calculated the parameters of the open charm-bottom
axial-vector tetraquark states $Z_q$ and $Z_{s}$ within QCD sum rule method.
Their masses and meson-current couplings  have been obtained using the
two-point sum rule method. In these calculations for $Z_q$ and $Z_s$ we have used
the symmetric in color indices interpolating currents by assuming that they
are ground states in corresponding tetraquark multiplets. Indeed, one can anticipate
that $Z_q$ and $Z_s$ are the axial-vector components of the $1S$
diquark-antidiquark  $[cq][\bar {b} \bar {q} ]$ and
$[cs][\bar {b} \bar {s}]$ multiplets, respectively.

During last years some progress was archived in investigation of the
$[cq][\bar{c}\bar{q}]$ and $[cs][\bar{c}\bar{s}]$ multiplets, and
classification of the observed hidden-charm tetraquarks as their possible members
(see, Refs.\ \cite{Maiani:2014,Maiani:2016wlq}). Thus,  within the "type-II" model
elaborated in these works, the authors not only identified the multiplet levels with
discovered tetraquarks, but also estimated masses of the states, which had not yet been  observed.
This model is founded on some assumptions about a nature of inter-quark and inter-diquark
interactions, and considers spin-spin interactions within diquarks as decisive source of
splitting inside of the multiplet.

The information useful for our purposes is accumulated in the axial-vector sector of
these multiplets. The axial-vector $J^{PC}=1^{++}$  particle in the ground-state
$[cq][\bar{c}\bar{q}]$ multiplet was identified with the well-known $X(3872)$ resonance.
The similar analysis carried out for the multiplet of $[cs][\bar {c} \bar s]$ states
demonstrated that its $J^{PC}=1^{++}$ level may  be considered as $X(4140)$. The mass difference
of the axial-vector resonances  belonging to $"q"$ and $"s"$ hidden-charm multiplets is
\begin{equation}
X(4140)-X(3872)\approx 270\,  \mathrm{MeV}.
\label{eq:Mult}
\end{equation}
In the present work we have evaluated masses of the axial-vector states from the
$[cq][\bar {b} \bar{q} ]$ and $[cs][\bar {b} \bar s]$ multiplets. The mass shift between these
multiplets
\begin{equation}
m_{Z_s}-m_{Z_q}\approx 240\,  \mathrm{MeV},
\end{equation}
is in  nice agreement with Eq.\ (\ref{eq:Mult}).

Another question to be addressed here is connected with masses of excited states, which
in sum rule calculations determine continuum threshold $s_0$.
We have found that for $[cq][\bar {b} \bar{q} ]$ and $[cs][\bar {b} \bar{s}]$
multiplets sum rule calculations fix the lower bounds of the parameter
$s_0$  as $s_0=59\,\, \mathrm{GeV^2}$ and $s_0=60\,\, \mathrm{GeV^2}$, respectively.
This means that sum rule has placed a first excited
state to position  $\sqrt s_0$. In order to estimate a gap between the excited and ground
states we invoke $\sqrt s_0$ and  central values of $Z_q$ and $Z_s$ masses. Then,
it is not difficult to see, that for
$[cq][\bar {b}\bar{q}]$ type tetraquarks, it equals to
\begin{equation}
\sqrt s_0\, \mathrm{GeV}-7.06\, \mathrm{GeV}\approx 0.62\, \mathrm{GeV},
\label{eq:EStates1}
\end{equation}
whereas for the $[cs][\bar {b}\bar{s}]$ one gets
\begin{equation}
\sqrt s_0\, \mathrm{GeV}-7.30\, \mathrm{GeV}\approx 0.45\, \mathrm{GeV}.
\label{eq:EStates2}
\end{equation}

The masses of $1S$ and $2S$ states with  $J^{PC}=1^{+-}$ from the $[cq][\bar{c} \bar{q}^{\prime}]$
multiplet were calculated  by means of the two-point sum rule method in Ref.\ \cite{Wang:2014vha}.
The ground-state  level $1S$ was identified with the resonance $Z_{c}(3900)$, whereas the resonance
$Z(4430)$ was included into a multiplet of the excited $2S$ states. If this assignment is correct,
then the experimental data provides the mass difference between the ground and first radially excited
states, which is equal to $530\, \mathrm{MeV}$. Results of the calculations led to predictions $M_{Z_c(3900)}=3.91^{+21}_{-17}\, \mathrm{GeV}$ and $M_{Z_c(4430)}=4.51^{+17}_{-09}\, \mathrm{GeV}$, and to the mass difference $\sim 600\, \mathrm{MeV}$.

The  $1S$ and $2S$ multiplets of $[cs][\bar{c} \bar{s}]$ tetraquarks were explored in the context of
the "type-II" model in Ref.\ \cite{Maiani:2016wlq}. For the axial-vector levels $J^{PC}=1^{++}$
named there as $X$ states, the  $2S-1S$ gap is $4600\, \mathrm{MeV}-4140\, \mathrm{MeV}=460 \, \mathrm{MeV}$,
and for the particles $X^{(1)}$ and $X^{(2)}$ with the quantum numbers $J^{PC}=1^{+-}$
one gets $4600\, \mathrm{MeV}-4140\, \mathrm{MeV}=460 \, \mathrm{MeV}$ and
$4700\, \mathrm{MeV}-4274\, \mathrm{MeV}=426 \, \mathrm{MeV}$, respectively.
Comparison of these results with ones given by  Eqs.\ (\ref{eq:EStates1}) and (\ref{eq:EStates2})
can be considered  as confirmation of a self-consistent character of the performed analysis.

In the framework of QCD two-point sum rule approach masses of the open charm-bottom
diquark-antidiquark states  were previously calculated in Ref.\ \cite{Chen:2013aba}. For
masses of the axial-vector tetqaruarks $Z_{q}$ and $Z_{s}$ the authors found:
\begin{equation}
m_{Z_{q}}=7.10\pm 0.09\pm 0.06\pm 0.01\,\,\mathrm{GeV},
\end{equation}%
and
\begin{equation}
m_{Z_{s}}=7.11\pm 0.08\pm 0.05\pm 0.03\,\,\mathrm{GeV}.
\label{eq:Chen}
\end{equation}%
These predictions were extracted by using  the parameter $s_0=(55\pm 2)\,\,\mathrm{GeV}^2$ in
calculations of $m_{Z_q}$ and $m_{Z_s}$, and $M^2=(7.9-8.2)\,\,\mathrm{GeV}^2$ and
$M^2=(6.7-7.9)\,\,\mathrm{GeV}^2$ for $"q"$ and $"s"$ states, respectively. It is seen, that
mass differences $m_{Z_{s}}-m_{Z_{q}}\approx 10 \,\,\mathrm{MeV}$
and $\sqrt s_0-m_{Z_q}\approx \sqrt s_0-m_{Z_s}\approx 180 \,\,\mathrm{MeV}$ can
be neither   included into $"q"-"s"$  mass-hierarchy scheme of the ground state
tetraquarks nor accepted as giving correct mass shift between $1S$ and $2S$ multiplets.
Our results for $m_{Z_q}$ and $m_{Z_s}$, if  differences are ignored in chosen windows for the parameters
$s_0$ and $M^2$, within theoretical errors may be considered as being in agreement with the
predictions of Ref.\  \cite{Chen:2013aba}. But in our case the central value of $m_{Z_s}$
allows the decay process $Z_s \to B_c \phi$, whereas for $m_{Z_s}$ from Eq.\ (\ref{eq:Chen})
it remains among kinematically forbidden channels.

We have also calculated the widths of the  $Z_{q}\to B_{c}\rho $  and $Z_{s}\to B_{c}\phi$
decays, which are new results of this work. Obtained predictions for
$\Gamma(Z_{q}\to B_{c}\rho) $  and $\Gamma(Z_{s}\to B_{c}\phi)$ show that $Z_q$ may be considered as a narrow resonance, whereas $Z_s$ belongs to a class of wide tetraquark states.

Investigation of the open charm-bottom axial-vector tetraquarks performed
in the present work within the diquark-antidiquark picture led to quite interesting
predictions. Theoretical explorations of other members of the $[cq][\bar {b} \bar{q} ]$
and $[cs][\bar {b} \bar{s}]$ tetraquark multiplets, as well as their experimental studies
may shed light on the nature of multi-quark hadrons.

\section*{ACKNOWLEDGEMENTS}

Work of K.~A. was financed by TUBITAK under the grant No. 115F183.

\end{document}